# Driving energetically-unfavorable dehydrogenation dynamics with plasmonics

Katherine Sytwu[1], Michal Vadai[2], Fariah Hayee[3], Daniel K. Angell[2], Alan Dai[4], Jefferson Dixon[5], Jennifer A. Dionne[2]*

[1]Department of Applied Physics, Stanford University, 348 Via Pueblo, Stanford, CA 94305

[2]Department of Materials Science and Engineering, Stanford University, 496 Lomita Mall, Stanford, CA 94305

[3]Department of Electrical Engineering, Stanford University, 350 Jane Stanford Way, Stanford, CA 94305

[4]Department of Chemical Engineering, Stanford University, 443 Via Ortega, Stanford, CA 94305

[5]Department of Mechanical Engineering, Stanford University, 440 Escondido Mall, Stanford, CA 94305

*Correspondence to: jdionne@stanford.edu

**Abstract**

Nanoparticle surface structure and geometry generally dictate where chemical transformations occur, with the low-coordination-number, high-radius-of-curvature sites being energetically-preferred. Here, we show how optical excitation of plasmons enables spatially-controlled chemical transformations, including access to sites which, without illumination, would be energetically-unfavorable. We design a crossed-bar Au-$PdH_x$ antenna-reactor system that localizes electromagnetic enhancement away from the innately reactive $PdH_x$ nanorod tips. Using optically-coupled *in situ* environmental transmission electron microscopy, we track the dehydrogenation of individual antenna-reactor pairs with varying optical illumination intensity, wavelength, and hydrogen pressure. Our *in situ* experiments show that plasmons enable new catalytic sites, including hydrogenation dissociation at the nanorod faces. Molecular dynamics simulations confirm that these new nucleation sites are energetically unfavorable in equilibrium and only accessible via tailored plasmonic excitation.

**Main Text**

Nanoparticle transformations and their transient states drive the successes and failures of many chemical processes. In catalysis, for example, nanoparticle catalysts undergo structural and compositional rearrangements depending on environmental factors like temperature and chemical composition *(1,2)*. These transient states support distinct electronic and material configurations from equilibrium and are not only crucial intermediate steps for material transformations like alloying *(3)*, but are also attributed to increases and decreases in product selectivity for a variety of reactions, including $CO_2$ hydrogenation *(4)*, CO oxidation *(5)*, and methanol steam reformation *(6)*. Control over transient material transformations could not only avoid reaction pathways that are less effective, but also create new electronic or structural states that better traverse the chemical reaction space. However, modifying these transformation dynamics is an outstanding challenge, often requiring atomic-level design and control. Notably, there is a mismatch in length-scale between the atomic-scale structural features (i.e. atomic coordination number, surface strain, etc.) which influence transformation dynamics and the extrinsic parameters (i.e. temperature, chemical composition, chemical environment, etc.) that can be controlled.

Fortuitously, recent advances in the field of plasmonics offer a solution as to how this size-mismatch can be bridged. Optical excitation of localized surface plasmon resonances (LSPRs) can nanoscopically confine light and control certain extrinsic properties like the lattice and electronic temperature across a nanoparticle with subparticle resolution *(7)*. Regions of high electromagnetic field intensity, commonly referred to as electromagnetic (EM) "hot spots," underpin increased chemical kinetics seen in plasmon photocatalysis and plasmon-driven nanoparticle growth *(8-10)*; they can also modify electronic and molecular energy levels and access excited state dynamics, enabling reaction pathways that are difficult or impossible to achieve with typical conditions (*11*). As the spatial distribution of electromagnetic hot spots is determined by nanoparticle geometry, LSPRs could provide the required nanoscale spatial control over transient states, reshaping the energy landscape of reactants, intermediates, and products *(12)*.

Here, we provide a proof-of-concept demonstration that plasmons can enable new transient states in chemical transformations. In particular, we show that plasmons can transform a normally non-reactive nanoparticle surface facet into the preferred reaction site. Such nanoscale control allows, in principle, for the entire nanoparticle surface to catalyze reactions rather than just the active sites, meaning that chemical activity is not solely defined by structural properties. This is particularly important for catalysts operating at lower temperatures, for instance in electrochemistry, where the role of active sites are more prominent *(13)*. We also note that this result is distinct from most other plasmon photocatalysis studies, where the regions of highest EM enhancement correspond with the regions of high chemical activity without illumination *(12)*.

To demonstrate this result, it is crucial to simultaneously achieve 1) sub-diffraction spatial information to resolve different features across a nanocatalyst, 2) sufficient temporal resolution to identify transient events, and 3) chemical/structural information to distinguish chemical and/or physical transformations. Though there has been much progress in techniques like super-resolution chemical imaging *(14)*, ex-situ nanoparticle markers *(8)*, environmental electron spectroscopy *(9)*, and *in situ* surface or tip-enhanced Raman spectroscopy *(15)*, very few can achieve simultaneous spatial, temporal, and chemical information at the nanoscale. We do so here by utilizing optically-coupled environmental transmission electron microscopy (TEM) *(16)*, as shown in Figure 1A. TEM provides simultaneous imaging of the entire nanoparticle with ~100ms temporal resolution. The additional environmental capability allows us to control the partial pressure of the gas environment around our sample and track our reaction *in situ*.

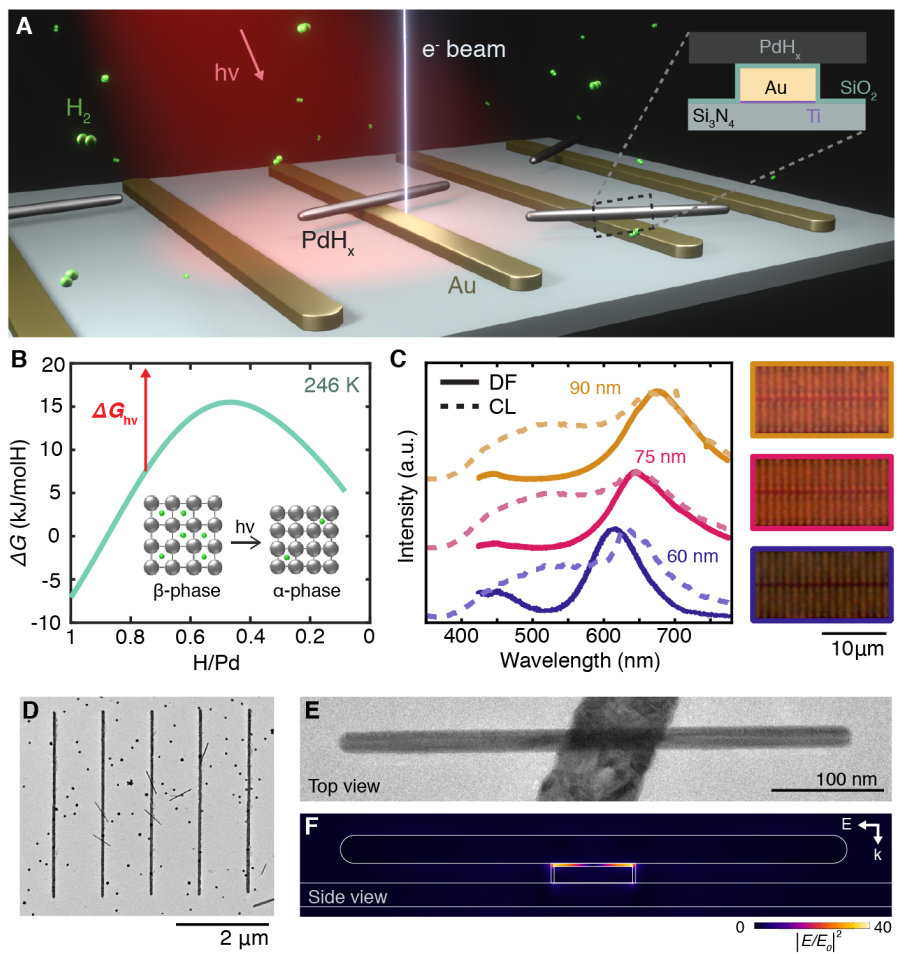

**Fig. 1.** The plasmonic Au-PdH$_x$ crossed-bar nanostructure. (A) Schematic of the experiment in real space with the inset showing a cross-sectional view of the intersection between the Au nanobar and PdH$_x$ nanorod. (B) Schematic of the experiment in energy space, showing the Gibbs free energy of formation for PdH$_x$ at 246K, calculated using a mean-field lattice gas model with finite size corrections (see Supplementary Materials). The incoming light/plasmon provides additional energy, which is enough to overcome the activation energy barrier to push the PdH$_x$ nanorod from the thermodynamically-stable β-phase to the metastable α-phase. (C) Dark field (DF) and cathodoluminescence (CL) spectra of SiO$_2$ coated Au nanobars of varying width (60-90nm) and their corresponding dark field images (right). Note that the CL spectra includes Pd nanoparticles scattered on top of the sample. For the 60nm wide Au nanobars, the CL spectra shows a peak at 635nm, red-shifted from the LSPR at 610nm, due to e-beam induced radiative defects in the SiO$_2$ layer which are further enhanced and convoluted with the LSPR response (see Supplementary Materials). (D) A wide-field view TEM image of five Au nanobar structures with Pd nanorods and other Pd nanoparticles randomly distributed. (E) TEM image showing a top-down view of a single Au-Pd crossed-bar nanoparticle system. (F) Cross-sectional view of the simulated electromagnetic enhancement of the Au-Pd system on resonance under transverse excitation. Simulation was done using a commercial finite element method solver (COMSOL).

Our model reaction is the dehydrogenation of palladium hydride ($PdH_x$) which consists of hydrogen dissociation at the Pd surface followed by interstitial hydrogen desorbing from the Pd matrix, causing a phase change from a hydrogen-rich β-phase (x~0.6) to a hydrogen-poor α-phase (x~0.01). Pd nanoparticles are commonly used as catalysts for hydrogenation reactions, but their subsequent transformation into palladium hydride can be detrimental for catalytic activity and selectivity *(17)*. Our previous studies of hydrogenation in $PdH_x$ nanorods and other nanoparticles have shown that the new phase exclusively nucleates at corners or tips of nanoparticles *(18,19)*. This dissociation phase transition is controlled by the surrounding temperature and chemical potential (i.e. $H_2$ pressure) and at 246K is endothermic (ΔH>0) and non-spontaneous (ΔG >>$k_b T$). As illustrated in Figure 1B, we use light via LSPRs to provide enough energy to the system such that it overcomes the energetic activation barrier to go from the thermodynamically-stable β-phase to the metastable α-phase. Experimental studies have reported this activation barrier to span from 20 to 80 kJ/molH (0.2-0.8eV) *(20,21)*, which can be readily overcome by visible-frequency photons.

Our plasmonic antenna-reactor structure consists of a 60-90nm wide plasmonic Au nanobar antenna surrounded by a thin 2nm $SiO_2$ spacer layer and a $PdH_x$ nanorod crossed on top. Importantly, this geometry spatially separates the EM hot spots from the favorable dissociation sites at the nanorod tips. To fabricate this structure, we lithographically pattern Au nanobars onto a 30nm thick $Si_3N_4$ TEM grid and then use atomic layer deposition to evenly coat the sample with $SiO_2$ to create a spacer layer (see Methods). These plasmonic antennas support a transverse-mode LSPR whose resonant wavelength ranges from 600-680nm with varying nanobar width, as seen in the dark field images and spectra in Fig 1C and further verified via FDTD simulations (Figure S10). Next, colloidally synthesized pentatwinned Pd nanorods of 350-550nm in length are dropcasted onto the prefabricated TEM grid and the crossed geometry is formed by random alignment, as seen in Figures 1D and 1E. Cathodoluminescence (CL) spectra of the structure with the dropcasted Pd nanoparticles reveal a LSPR peak that agrees with the dark field spectra (note that the broad shoulder from 400-600nm originates from the $Si_3N_4$ background). At the crossing junction, EM hot spots are formed in the gap between the Pd nanorod and Au nanobar (Figure 1F). Therefore, the LSPR enhancement spatially overlaps with a region of the $PdH_x$ nanorod that is not favorable for hydrogen dissociation.

To track the dehydrogenation, we use displaced-aperture dark field (DADF) imaging to exploit the ~2-3% decrease in lattice parameter as the particle goes from β- to α- phase *(22)*. After setting the temperature and hydrogen pressure to such that the nanoparticle is stable in the hydrogenated β-phase (see Methods), we judiciously place an objective aperture around a diffraction point of the $PdH_x$ nanorod (Figure 2A) to obtain a dark field image of the bottom $PdH_x$ nanorod crystallite (Figure 2B, top). We note that the weaker signal at the intersection of the two nanostructures is due to the strong scattering of electrons off the Au nanobar. Upon an external stimuli (i.e. a drop in pressure or illumination), the particle desorbs $H_2$ and transitions to the dehydrogenated α-phase; this lattice shrinkage results in the diffraction pattern expanding, which shifts the diffraction point outside the objective aperture. Under these new conditions, the dark field image cannot be formed and the image disappears (Figure 2B, bottom). Therefore, the dark field image is a proxy for the hydrogenation state of the particle, with signal indicating β-phase and lack of signal indicating α-phase, providing real-time information on the phase state of the nanoparticle with nanometer spatial resolution.

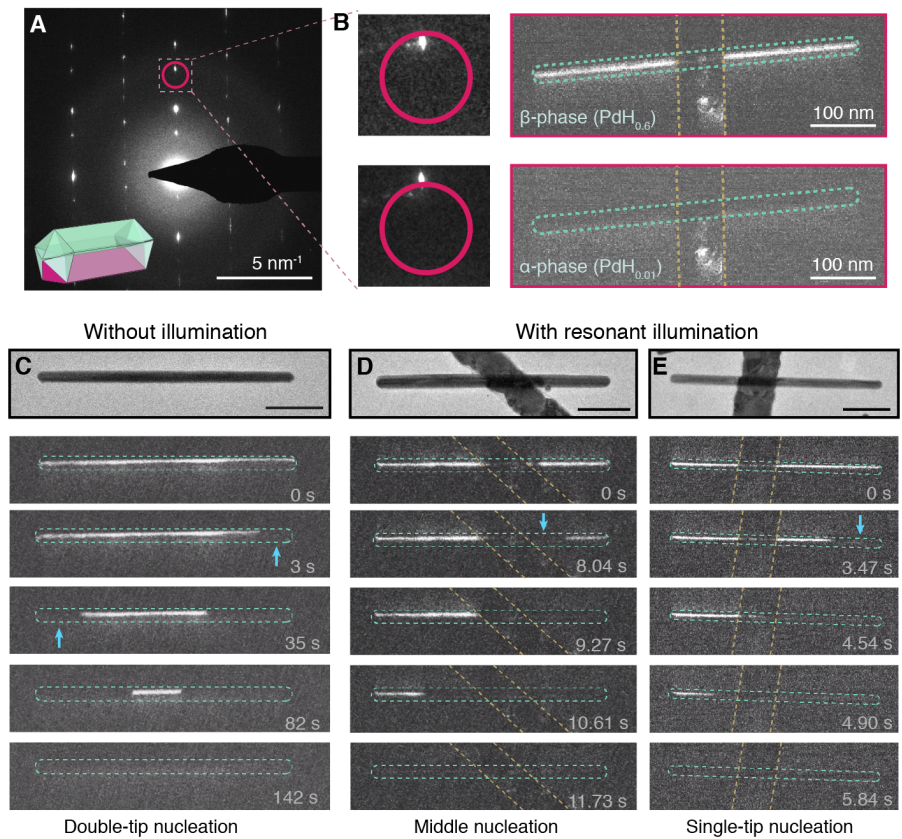

**Fig. 2.** Live Displaced-Aperture Dark Field (DADF) imaging of the $PdH_x$ dissociation dynamics. (A) Electron diffraction pattern of a single pentatwinned $PdH_x$ nanorod. The pink circle designates the approximate location of the objective aperture around the (200) diffraction point for DADF imaging, with inset schematic showing the bottom crystallite to which it belongs. (B) Zoom-in of the diffraction points relative to the objective aperture (pink) and corresponding dark field image of the nanorod in β-phase (top) and after the transformation into α-phase (bottom). Only the $PdH_x$ diffraction point moves; the diffraction point(s) corresponding to the Au crystallite remains within the objective aperture and so the dark field image of the Au crystallite remains in both images. (C) TEM image and DADF snapshots of the dehydrogenation mechanism of a 498nm $PdH_x$ nanorod without illumination, with blue arrows pointing out nucleation sites. Time labels refer to the time from the first nucleation event. See supplementary for more examples. (D,E) TEM and DADF snapshots of the dehydrogenation mechanism of the Au- $PdH_x$ system under resonant illumination ((D) 627nm with 50nm bandwidth and (E) 640nm with 20nm bandwidth) for a (D) 446nm and (E) 447nm nanorod. Under resonant illumination, $PdH_x$ nanorods show two types of behavior: (D) middle nucleation and (E) single-tip nucleation. Time labels refer to the time from the start of illumination. Aqua and yellow dashed lines serve as guides-to-the-eye to denote nanoparticle boundaries. All scale bars are 100nm.

Using this DADF technique, we first confirm that $PdH_x$ nanorods naturally dehydrogenate at their tips. Figure 2C shows a series of dark field images of an isolated $PdH_x$ nanorod as it naturally dehydrogenates following continual drops in hydrogen pressure (see Methods). In Fig 2C, we see that the signal for the right nanorod tip starts to disappear, implying that this tip is now in the α-phase. 35 seconds after the phase transition began, the left tip starts to disappear as well, indicating a second nucleation from the opposite tip of the nanorod, until finally, the entire image has disappeared and the particle has fully transformed into the α-phase. In all of our measurements of isolated $PdH_x$ nanorods (8 measurements over 5 $PdH_x$ nanorods), we exclusively observe nucleation of the new α-phase from both ends of the nanoparticle, herein referred to as double tip nucleation. This is consistent with our prior measurements of the thermodynamic hydrogenation behavior of long nanorods, which show that at these Pd nanorod lengths (>350nm), nanorods are more likely to nucleate their new phase at both tips *(19)*. Therefore, not only are the tips the "active site" for our dissolution process, but $PdH_x$ nanorods of this length energetically prefer having two nucleation sites.

The addition of plasmonic hot spots modifies this behavior. We set the $H_2$ partial pressure such that the nanorods are kinetically-trapped in the β-phase and unlikely to switch to the α-phase. Using a tunable pulsed laser, we illuminate our sample near its resonant wavelength and track the behavior of a single crossed-bar system. We observe two distinct behaviors, both different from the case of an isolated $PdH_x$ nanorod: for some nanorods, the middle section of the $PdH_x$ nanorod dehydrogenates first, near the electromagnetic hot spot (Figure 2C). One phase-front then propagates towards the nearest nanorod tip, and then the opposite phase-front propagates towards the other end. Besides middle nucleation, we also observe nanorods that, upon illumination, dehydrogenate from exclusively *one* of the tips; here, unlike the dark condition, the new phase propagates to the other end without a second nucleation event at the opposite tip (Figure 2D). Both behaviors under illumination are notably different from the $PdH_x$ nanorod's behavior without plasmonic enhancement, which exclusively nucleates from both ends of the nanorod.

Due to differences in experimental setup, we cannot make quantitative comparisons between the kinetics with and without illumination, but we can qualitatively note that the plasmon-driven dissociation process is faster than the without illumination case, in line with our prior observations of $PdH_x$ nanocubes *(16)*. Furthermore, upon turning illumination off, some observed particles immediately switch back to β-phase, indicating that this phase transition is indeed driven by the optical illumination rather than just electron-beam excitation (see Supplementary Materials).

Wavelength-dependent studies further verify this is a plasmon-driven process. We track the nucleation site in two individual Au-$PdH_x$ pairs as a function of illumination wavelength and power to confirm LSPR dependence. By comparing electron energy loss spectroscopy (EELS) measurements with wavelength-dependent nucleation dynamics of the same nanobar system, we find that middle nucleation only occurs for illumination wavelengths that overlap with the plasmon resonance of the crossed-bar structure. As an example, in Fig 3A, we plot the LSPR mode and the corresponding EELS map of a Au-Pd nanostructure, showing that the transverse plasmon mode peaks at 2.07eV (599nm) and is spatially localized to the middle of the Pd nanorod. When the system is then illuminated, the nanorod only undergoes middle nucleation under illumination from 575-675nm, or around the LSPR peak (Figure 3B). This is also consistent with observed faster reaction times under 600nm center illumination (see Supplementary Materials). Notably, illumination at higher energies (lower wavelengths) but same power does not induce middle nucleation, confirming that resonance conditions are needed and this new behavior is not driven by an absorption process originating from the substrate (i.e. the radiative defect in $SiO_2$). Likewise, at lower illumination powers, we do not observe nucleation of the dehydrogenated phase, even at the resonant illumination wavelength, suggesting that we have to overcome a 'threshold' power in order to induce the phase transition. In Fig 3C and 3D, we see similar behavior for a nanostructure whose LSPR is extracted to be at 1.98eV (626nm). This system demonstrates middle nucleation at a wider range of illumination wavelengths, likely due to the higher average illumination power. Here, we also observe that off-resonant illumination at 700-725nm can induce a phase transition but only at higher illumination powers than the on-resonant illumination cases around 600-650nm.

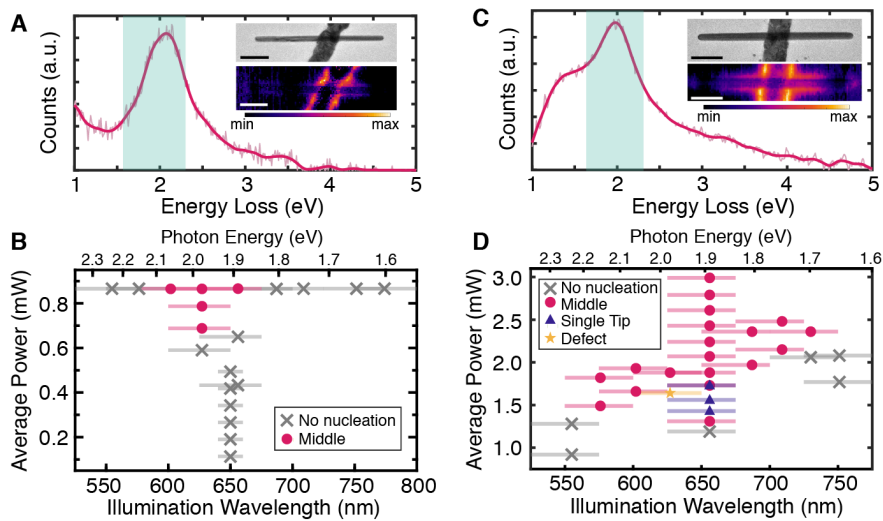

**Fig. 3.** Illumination wavelength and power dependence of nucleation site of two individual Au-$PdH_x$ pairs. Electron energy loss spectra (A,C) of the transverse plasmon mode (pink is smoothed spectra), with the corresponding TEM image and EELS map of the nanoparticle as an inset. Blue shaded region indicates the illumination range in the below nucleation site experiments. All scale bars are 100nm. (B,D) Nucleation site of the α-phase for various illumination wavelengths and powers on a single Au-$PdH_x$ pair, with each data point summarizing a DADF image series (as shown in Figure 2). Bars show the illumination bandwidth. Measurements were performed in (B) 77Pa and (D) 84Pa of $H_2$ gas. (A) and (B) are performed on the same nanostructure; similarly for (B) and (D).

This plasmon-induced behavior not only depends on the LSPR characteristics but also the thermodynamic state of the $PdH_x$ nanorod, indicated by the surrounding hydrogen pressure. To isolate the role of the inherent thermodynamics of the $PdH_x$ nanorod, we track the nucleation site in 22 Au-$PdH_x$ pairs under constant resonant illumination at various pressures around the natural dehydrogenation pressure, as schematically depicted by the isotherm in Figure 4A. Similar to before, we hydrogenate the nanoparticles, and then set the pressure of interest, waiting 30min before illumination to verify that the nanorods are in a near-equilibrium or kinetically trapped hydrogenated state. As individual nanoparticles have different dehydrogenation pressures and illumination intensities (see Supplementary Materials), we can only compare across a single Au-PdH$_x$ pair, not across particles. Under illumination, we observe five different phase transition mechanisms: no phase transition, single-tip nucleation (13 particles), double-tip nucleation (1 particle), middle nucleation (9 particles) and defect nucleation (1 particle), the statistics of which are summarized in Figure 4B. Many particles showed different dehydrogenation mechanisms depending on the surrounding hydrogen pressure, six of which are highlighted in Figure 4C (more statistics in Figure S9). Consistently, at higher pressures (i.e. when the PdH$_x$ nanorod is very stably in its β-phase), the energetic barrier to nucleate a phase transition is too high such that we either observe no dehydrogenation within our data acquisition period (180s) or a more energetically favorable mechanism, like single-tip nucleation. However, once the pressure is lowered such that we are closer to the natural dehydrogenation pressure, we start to see both middle and single-tip nucleation.

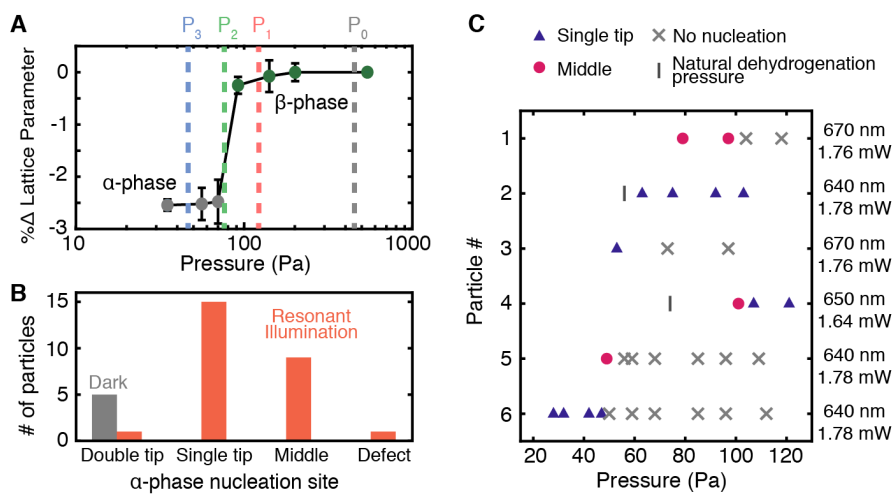

**Fig. 4.** Hydrogen pressure dependence of nucleation site. (A) Isotherm of a PdH$_x$ nanorod taken from diffraction patterns as it goes from β-phase to α-phase. Dotted lines schematically outline the pressure-dependent measurements; the environment is first set to P$_0$ to get the particle stably in β-phase, and then lowered to P$_1$, after which it is illuminated and the nucleation site is recorded. The process then repeats for pressures P$_2$, P$_3$, etc. (B) Particle statistics of α-phase nucleation site under dark and resonant illumination conditions for five particles under dark conditions and twenty-one particles under illumination conditions. Some particles are double counted since they show different behavior depending on the surrounding hydrogen pressure. (C) Pressure-dependent nucleation behavior for six different Au-PdH$_x$ pairs under resonant illumination (with average illumination wavelength and power specified in the right column). Additional particle statistics shown in Fig. S9.

From a macroscopic view, our results demonstrate how LSPRs not only accelerate reactions, but also enable new energetically-unfavorable chemical transformations that are tailored by plasmonic design. We classify the PdH$_x$ phase transition into three distinct behaviors: double-tip nucleation, single-tip nucleation, and middle/defect nucleation; and hypothesize that plasmons not only supply the system with enough energy to undergo a less favorable transition but careful tailoring of the conditions also allows for transition mechanisms never seen before (i.e. middle nucleation). Double-tip nucleation is seen in all of the dehydrogenation videos of isolated PdH$_x$ nanorods without illumination, suggesting that it is the most energetically-favorable phase transition mechanism for long PdH$_x$ nanorods. On the other hand, when paired with a plasmonic antenna, the majority of PdH$_x$ nanorods (20 out of 22) show either single-tip nucleation or middle/defect nucleation under illumination at low hydrogen pressures. We hypothesize that single-tip nucleation is the next favorable mechanism, as it is geometrically similar and possibly a kinetically faster version of double-tip nucleation. Finally, middle/defect nucleation seems to be the least favorable mechanism and requires careful tailoring of the illumination and hydrogen pressure environment.

To verify this hypothesis, we qualitatively compare the energies of our three nucleation configurations using molecular dynamics simulations with an embedded atom method (EAM) interatomic potential *(23)*. The EAM potential has been shown to correctly model and replicate the strain-induced thermodynamic miscibility gap for both nanoparticle and bulk PdH$_x$. To model our nanorod, we simulate three nanobars of 20nm width and various lengths with identical β-phase volume fractions but different spatial configurations, simulating double-tip, single-tip and middle nucleation. We then calculate the Gibbs free energy of each configuration (see Supplementary Materials) and compare it to that of a fully hydrogenated nanobar, or our starting configuration (Figure 5A). We find that our simulations exactly match our hypothesis. Double-tip nucleation is the closest in energy to the fully hydrogenated (β-phase) nanobar while middle nucleation has the largest energy difference, suggesting that double-tip nucleation would be the most energetically-favorable pathway while middle nucleation would be the least. Interestingly, single-tip and middle nucleation are closer in energy than the double-tip configuration, suggesting that we would see near-equal probabilities of either behavior. This also explains why we see both single-tip and middle nucleation under resonant illumination. Therefore, plasmon excitations can influence reaction pathways, even causing behavior that is otherwise energetically unfavorable.

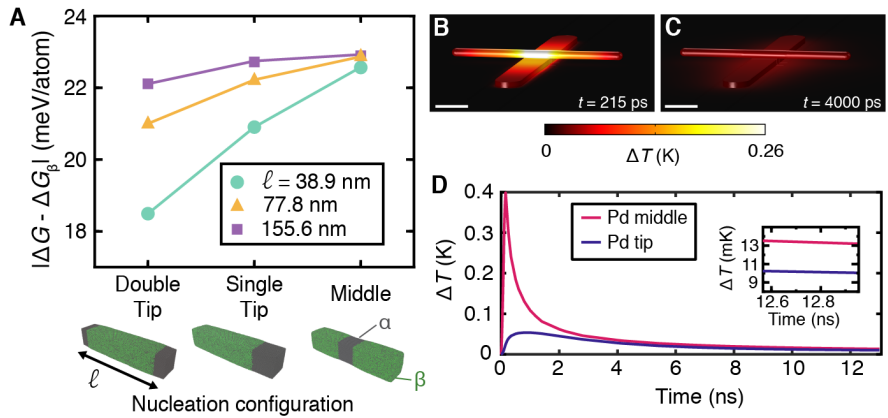

**Fig. 5.** Simulations of the Au-PdH$_x$ system. (A) The difference in free energy between three different nucleation configurations and a fully hydrogenated nanobar ($\Delta G_\beta$), calculated via molecular dynamics simulations of 20nm width and 38.9nm (aqua, circle), 77.8nm (yellow, triangle), and 155.6nm (purple, square) length nanobars. For all nanobar lengths, the double-tip

nucleation configuration has the closest energy to the fully hydrogenated nanobar. (B,C,D) Finite element modeling of plasmon-induced heating under pulsed illumination for the Au-Pd structure. Temperature distribution across the Au-Pd structure at 215 ps (B) and after it has some time to equilibrate at 4 ns (C). Scale bars are 100nm. Initial pulse of light is at 100ps. (D) The temperature at the Pd nanorod middle (pink) versus tip (blue) as a function of time after the pulse of light at 100ps. At 12.8ns (inset), the temperature difference between the middle and tip is 3.2mK.

From a microscopic view, the LSPR and its decay products of light, hot carriers, and heat induce hydrogen dissociation at the electromagnetic hot spot, locally reducing the nucleation energy barrier. The hot carriers' short lifetime (10's of fs) and mean free path (~40nm for Au and ~10-20nm for $PdH_x$ *(24)*) indicate that their spatial distribution should follow the localized profile of the EM enhancement. On the other hand, the thermal distribution, which persists on a longer timescale, initially follows the spatial profile of the EM enhancement (Figure 5B) but spatially broadens out over time due to nanoscale heat transfer (Figure 5C), suggesting that the single-tip nucleation process is driven by plasmonic heating. Middle nucleation, however, could be caused by a combination of all three decay products. The increased radiation at the plasmonic hot spot locally reduces the energy barrier for hydrogen recombination and desorption, allowing hot carriers to populate the necessary Pd-H orbitals, a phenomenon that explains our previous results of site-selectivity in $PdH_x$ nanocubes *(16, 25)*. However, unlike our previous $PdH_x$ nanocube system, here we potentially have non-uniform plasmonic heating across our $PdH_x$ nanorod reactor. We model the plasmon-induced heating and dissipation of the system under pulsed illumination, as shown in Fig 5D, and plot the time-dependent temperature of the Pd nanorod at two locations: in the middle of the nanorod and at one of the tips. Our simulations show that after 12.8ns, the time between sequential pulses, there remains a small but finite temperature difference between the middle and tip (3.2mK). Extrapolating this temperature difference to the ms timescale of our experiment requires unphysical linearization of the temporal response (see Supplementary Materials), leaving it inconclusive as to whether heating plays a dominant role or not. While identifying the mechanism is important in understanding how LSPRs affect the rate-limiting step, we note that the unique transformation mechanisms we observe can be achieved by any of the decay products; future work could utilize structures with more distinct thermal and non-thermal spatial distributions to further disentangle these mechanisms.

In summary, using *in-situ* environmental TEM combined with light illumination, we have demonstrated how plasmons modify the preferred active site of nanoparticles for dehydrogenation. Our proof-of-concept results demonstrate how both electromagnetic enhancement as well as innate material response dictate the subsequent plasmon-induced behavior, and how careful design of both can create new dynamics. By transforming non-reactive sites into the most favorable catalytic sites, plasmons could facilitate use of the entire catalytic surface for next-generation multiplexed photocatalysts. Beyond catalysis, such results might find utility in electro-optic and energy storage devices that rely on phase transformations, for example enabling site-specific metal-insulation transitions in $VO_2$ *(26)* or increased hydrogenation kinetics in $MgH_x$.